\documentclass[twocolumn,showkeys,aps,prb,showpacs]{revtex4-1}
\UseRawInputEncoding
\usepackage{graphicx}
\usepackage[CJKbookmarks,dvipdfm,colorlinks,linkcolor=red,citecolor=blue]{hyperref}

\begin{document}

\title{A possible way to achieve anomalous valley Hall effect  by piezoelectric effect in  $\mathrm{GdCl_2}$ monolayer}

\author{San-Dong Guo$^{1}$, Jing-Xin Zhu$^{1}$, Wen-Qi Mu$^{1}$  and  Bang-Gui Liu$^{2,3}$}
\affiliation{$^1$School of Electronic Engineering, Xi'an University of Posts and Telecommunications, Xi'an 710121, China}
\affiliation{$^2$ Beijing National Laboratory for Condensed Matter Physics, Institute of Physics, Chinese Academy of Sciences, Beijing 100190, People's Republic of China}
\affiliation{$^3$School of Physical Sciences, University of Chinese Academy of Sciences, Beijing 100190, People's Republic of China}
\begin{abstract}
Ferrovalley  materials can  achieve manipulation of the valley degree of freedom with  intrinsic spontaneous valley polarization introduced by their intrinsic ferromagnetism.  A good ferrovalley material should possess perpendicular magnetic anisotropy
(PMA), valence band maximum (VBM)/conduction band minimum (CBM) at valley points, strong ferromagnetic (FM) coupling and  proper valley splitting.
In this work,  the  monolayer $\mathrm{GdCl_2}$ is proposed  as a potential candidate
material for valleytronic applications by the first-principles calculations. It is proved that monolayer $\mathrm{GdCl_2}$ is a  FM semiconductor with the easy axis  along out of plane direction and strong FM coupling.  A
spontaneous valley polarization with a valley splitting
of 42.3 meV is produced  due to its intrinsic ferromagnetism and  spin orbital
coupling (SOC). Although the VBM of unstrained monolayer $\mathrm{GdCl_2}$ is away from valley points, a very small compressive strain (about 1\%) can make
VBM move to valley points. We  propose a possible way
to realize anomalous valley Hall effect  in monolayer $\mathrm{GdCl_2}$ by piezoelectric effect, not an  external
electric field, namely piezoelectric anomalous valley Hall effect (PAVHE).
This phenomenon could be classified as piezo-valleytronics,  being similar to piezotronics and piezophototronics.
The only independent piezoelectric  strain coefficient  $d_{11}$  is  -2.708 pm/V, which is  comparable to one of  classical bulk piezoelectric material $\alpha$-quartz ($d_{11}$=2.3 pm/V). The  biaxial in-plane strain and electronic correlation effects are considered to confirm
the reliability of our results. Finally, the monolayer $\mathrm{GdF_2}$  is predicted to be a ferrovalley  material with  dynamic and mechanical stabilities, PMA, VBM at valley points, strong FM coupling,  valley splitting of 47.6 meV,  and $d_{11}$ of 0.584 pm/V.
Our works  provide a possible way  to achieve anomalous valley Hall effect  by piezoelectric effect, which may stimulate further experimental works related with valleytronics.

\end{abstract}
\keywords{Valleytronics, Ferromagnetism, Piezoelectronics, 2D materials}

\pacs{71.20.-b, 77.65.-j, 72.15.Jf, 78.67.-n ~~~~~~~~~~~~~~~~~~~~~~~~~~~~~~~~~~~Email:sandongyuwang@163.com}

\maketitle

\section{Introduction}
 Rather than spin
and charge,  carriers in crystals
are also endowed with the valley degree of freedom, which is useful to
 process information and perform logic operations (valleytronics)\cite{q1,q2,q3,q4,q5,q6}.
  Two or more local energy extremes in
the conduction band or valence band, which are degenerate but inequivalent at
the inequivalent k points in the momentum space, are needed for a valley material.
 To realize applications of valleytronics, the   electrons or holes in different valleys must be selectively produced
or manipulated.  Although the possibility to achieve manipulation of the valley degree of freedom has  been proposed in certain  three-dimensional (3D)
materials\cite{q7},  the field of valleytronics is truly flourishing with the advent of two-dimensional (2D) materials.

\begin{figure}
  \includegraphics[width=8.0cm]{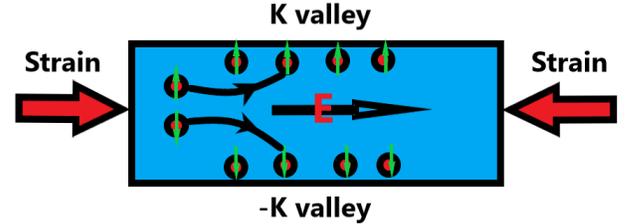}
  \caption{(Color online)Sketch of  anomalous valley Hall effect under an in-plane longitudinal electric field $E$, and the $E$ is induced with  uniaxial strain by piezoelectric effect.  Upward arrows and downward arrows  represent spin-up and spin-down carriers, respectively.
  Only one edge of the sample can accumulate the  charge carriers, and another edge will  accumulate ones, when reversing the
magnetization orientation. }\label{v-0}
\end{figure}

The reduction in dimensionality of 2D
materials results in
that space inversion symmetry
is often eliminated in 2D structures,
allowing  these materials to become piezoelectric\cite{q8}, which is also very important for valleytronics described by Berry curvature $\Omega(k)$.
In 2D hexagonal systems with broken
space inversion symmetry,  the Berry curvature  in the K and
-K valleys will be nonzero  along the
out of plane direction, and the Berry curvatures of two
valleys are  in opposite signs. If the time reversal
symmetry is also broken, their absolute
values are no longer identical, and the valley contrasting feature will
be induced.  Under an in-plane longitudinal electric field $E$,
the Bloch electrons in these 2D systems will  acquire an anomalous Hall velocity
$\upsilon$  due to $\upsilon\sim E\times\Omega(k)$\cite{q9}, and then the anomalous valley Hall effect will be produced, which can be achieved in ferrovalley  materials\cite{q10}.
Many  ferrovalley
materials have been predicted, such as  2H-$\mathrm{VSe_2}$\cite{q10}, $\mathrm{CrSi_2X_4}$ (X=N and P)\cite{q11}, $\mathrm{VAgP_2Se_6}$\cite{q12}, $\mathrm{LaBr_2}$\cite{q13,q13-1}, $\mathrm{VSi_2P_4}$\cite{q14},  $\mathrm{NbX_2}$ (X =S and Se)\cite{q15},
$\mathrm{Nb_3I_8}$\cite{q16}, $\mathrm{TiVI_6}$\cite{q17}.
It is a natural idea to induce in-plane longitudinal electric field $E$ with an  applied   uniaxial in-plane strain by piezoelectric effect, and then
anomalous valley Hall effect can be produced, which is
illustrated in \autoref{v-0}.

To well  achieve  PAVHE,  a  2D material should possess the strong FM coupling with PMA (The out-of-plane magnetization easy axis is important not
only for FM order but also for valley behavior.), the  appropriate energy band gap and valley splitting (The band gap and  valley splitting should be
large enough to overcome the thermal noise.), and
the pure in-plane piezoelectric effect  with only $d_{11}$ (The only independent $d_{11}$ means only in-plane longitudinal electric field.).
Recently, a kind of exotic 2D ferromagnetic semiconductors  $\mathrm{GdX_2}$ (X=Cl, Br and I)  based on rare-earth ions with f-electrons are predicted to have a large magnetization with high  Curie temperature beyond 220 K\cite{g1,g2}.
The monolayer $\mathrm{GdI_2}$ is predicted as  a promising candidate
material for valleytronic applications, which  is spontaneously
valley polarized with a giant splitting of 149 meV\cite{g3}.  However, $\mathrm{GdI_2}$  possesses in-plane magnetic anisotropy, not PMA\cite{g1,g2}.
Among $\mathrm{GdX_2}$ (X=Cl, Br and I) monolayers,   the
easy axis of only monolayer $\mathrm{GdCl_2}$ is along the out of plane direction\cite{g2}. The monolayer $\mathrm{GdCl_2}$  has $p\bar{6}m2$  point-group symmetry, which means that  only  independent $d_{11}$   is  nonzero.
\begin{figure}
  \includegraphics[width=6.0cm]{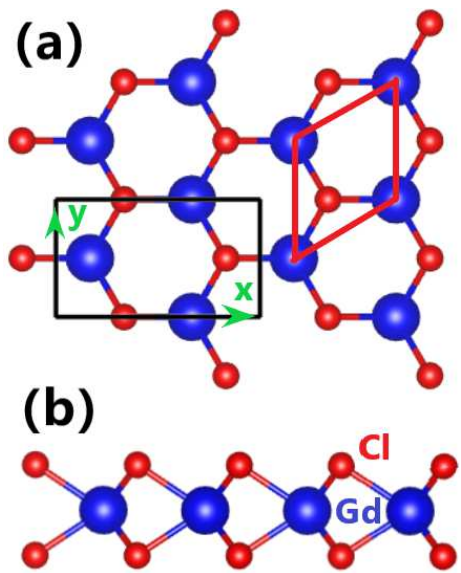}
  \caption{(Color online)The (a) top view and (b) side view of  crystal structure of  monolayer  $\mathrm{GdCl_2}$. The red and black frames represent the rhombus primitive cell  and  rectangle supercell. The  rectangle supercell is used to calculate the piezoelectric   stress  coefficients, whose width and height are defined as x and y directions, respectively.}\label{st}
\end{figure}

In light of PMA and independent $d_{11}$,  monolayer $\mathrm{GdCl_2}$ is likely to be a potential
ferrovalley material to realize PAVHE.  In this work,   we investigate the valley physics and  piezoelectric properties of monolayer $\mathrm{GdCl_2}$ by the first-principles calculations.  The  monolayer $\mathrm{GdCl_2}$ exhibits a pair of valleys in the valance band at
the K and -K points with a valley splitting
of 42.3 meV   due to its intrinsic ferromagnetism and  SOC. The predicted   $d_{11}$  is  -2.708 pm/V, which is  comparable to one of  $\alpha$-quartz ($d_{11}$=2.3 pm/V).  To confirm
the reliability of our results, the  biaxial in-plane strain and electronic correlation effects on  valley physics and  piezoelectric properties are considered. Finally, the monolayer $\mathrm{GdF_2}$  is predicted to be likely to be a potential
ferrovalley material.   Our works
provide  potential 2D valleytronic
materials to achieve PAVHE for developing high-performance and controllable valleytronics.

\begin{figure}
  \includegraphics[width=8cm]{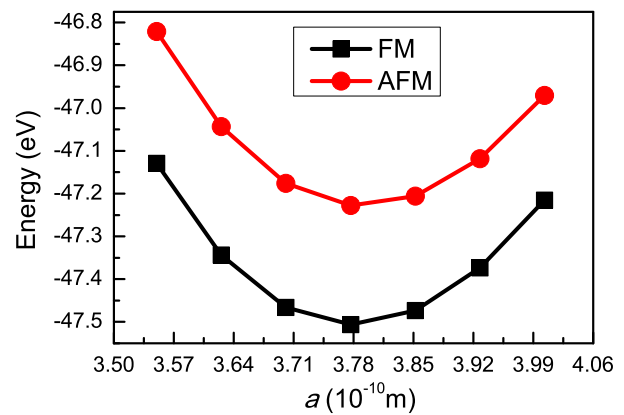}
  \caption{(Color online) Calculated energy of   AFM state and   FM state of monolayer  $\mathrm{GdCl_2}$ as a function of  lattice constants $a$  with rectangle supercell.  }\label{energy}
\end{figure}

\begin{figure*}
  \includegraphics[width=12cm]{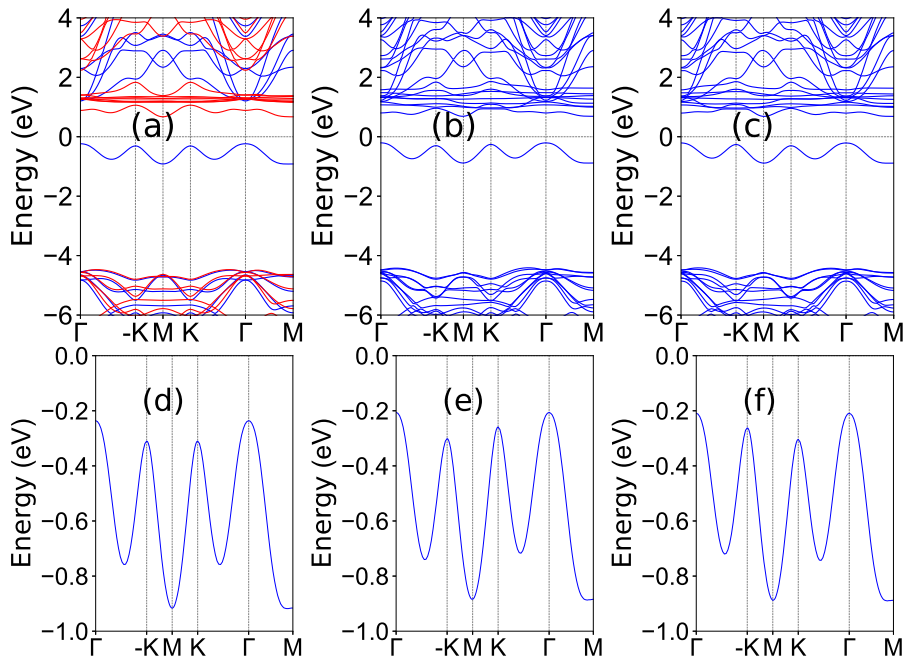}
  \caption{(Color online) The band structure of  monolayer  $\mathrm{GdCl_2}$ (a) without SOC; (b) and (c) with SOC for magnetic moment of Gd along the positive and negative z direction (out of plane), respectively.  The (d), (e) and (f) are  enlarged views of the valence bands near the Fermi level for (a), (b) and (c).}\label{band}
\end{figure*}

\begin{figure*}
  \includegraphics[width=15cm]{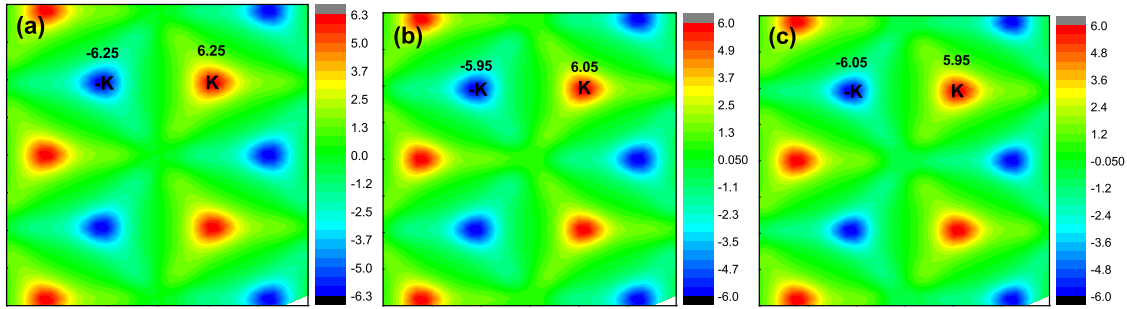}
  \caption{(Color online) Calculated Berry curvature distribution of monolayer  $\mathrm{GdCl_2}$  in the 2D Brillouin zone (a) without SOC; (b) and (c) with SOC for magnetic moment of Gd along the positive and negative z direction (out of plane), respectively.  }\label{berry}
\end{figure*}

\begin{figure*}
  \includegraphics[width=12cm]{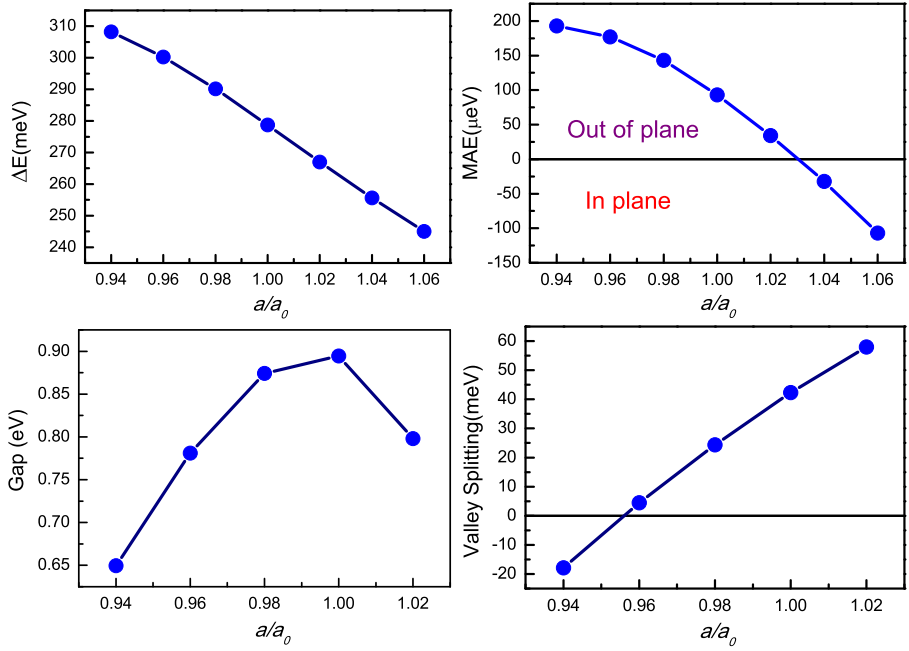}
\caption{(Color online)  (Top left) The energy difference between the AFM and FM states ($\Delta E$); (Top right) MAE; (Bottom left) energy band gap (Gap); (Bottom right) valley splitting
as a function of the applied biaxial strain $a/a_0$ for monolayer  $\mathrm{GdCl_2}$.}\label{s-1}
\end{figure*}

\begin{figure*}
  \includegraphics[width=16cm]{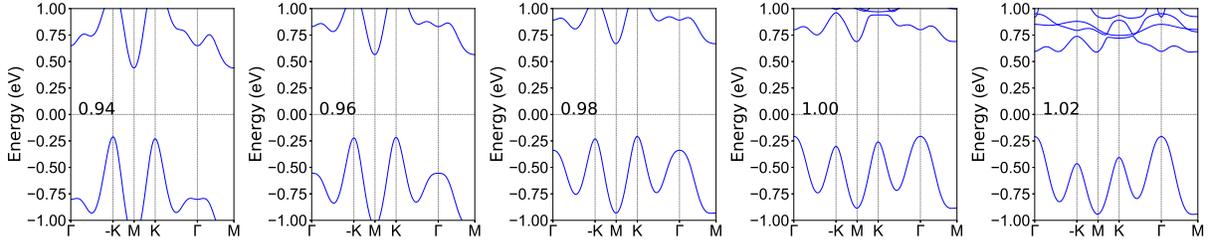}
\caption{(Color online) The energy band structures  of  monolayer  $\mathrm{GdCl_2}$   with $a/a_0$ from 0.94 to 1.02 by using GGA+SOC.}\label{s-band}
\end{figure*}

\begin{figure*}
  \includegraphics[width=15cm]{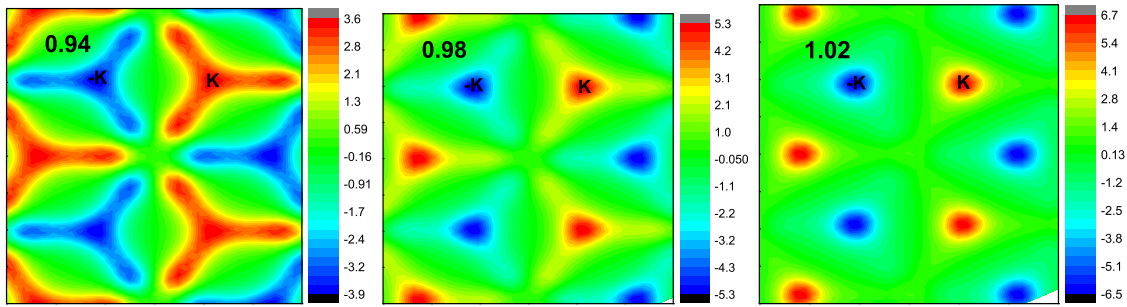}
\caption{(Color online) Calculated Berry curvature distribution of monolayer  $\mathrm{GdCl_2}$  in the 2D Brillouin zone  with $a/a_0$ being 0.94, 0.98 and 1.02 by using GGA+SOC.}\label{s-berry}
\end{figure*}
\begin{figure}
  \includegraphics[width=7cm]{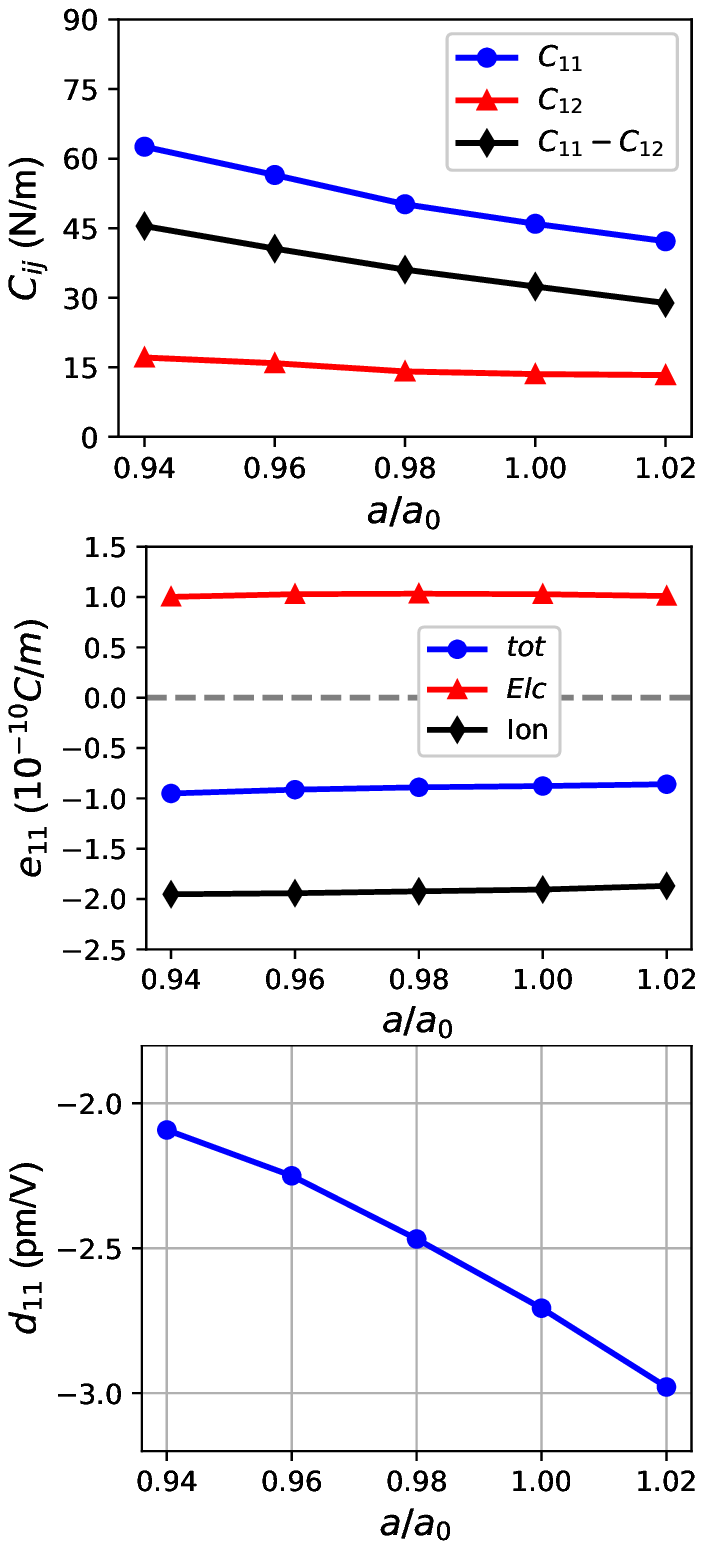}
\caption{(Color online) For monolayer  $\mathrm{GdCl_2}$, the elastic constants  $C_{ij}$,  the piezoelectric stress coefficient  ($e_{11}$) along with the ionic contribution and electronic contribution, and the piezoelectric strain coefficient ($d_{11}$)  with the application of biaxial strain (0.94 to 1.02).}\label{y-d}
\end{figure}

The rest of the paper is organized as follows. In the next
section, we shall give our computational details and methods.
 In  the next few sections,  we shall present structure and stability, electronic structure and valley Hall effect, and piezoelectric properties of monolayer  $\mathrm{GdCl_2}$, along with strain and electronic correlation effects on its valleytronic and piezoelectric properties. Finally, we shall give our discussion and conclusions.

\section{Computational detail}
First-principles calculations  with spin-polarization  are performed within density functional theory (DFT)\cite{1},  as implemented in the Vienna Ab
Initio Simulation Package (VASP)\cite{pv1,pv2,pv3} within the projector augmented-wave (PAW) method.  The generalized gradient
approximation (GGA) in the form of the Perdew-Burke-Ernzerhof (PBE) functional is used as the exchange-correlation interactions.
The kinetic energy cutoff is set to 500 eV, and   the total energy  convergence criterion  $10^{-8}$ eV is used.
The optimized  convergence criterion for atomic coordinates
 is less than 0.0001 $\mathrm{eV.{\AA}^{-1}}$  for force on each atom.
 The vacuum
space is set to more than 18 $\mathrm{{\AA}}$ to avoid adjacent interactions.
The 18$\times$18$\times$1 Monkhorst-Pack k-point mesh is used to sample the Brillouin zone for calculating electronic structures and elastic properties, and 10$\times$20$\times$1 Monkhorst-Pack k-point mesh for piezoelectric calculations.
To account for the localized nature of 4$f$ orbitals of Gd
atoms, a Hubbard correction $U_{eff}$  is employed within the
rotationally invariant approach proposed by Dudarev et al., where $U_{eff}$ is set as 4 eV, 5 eV, 8 eV\cite{g1,g2} for for monolayer $\mathrm{GdCl_2}$, $\mathrm{GdBr_2}$ and $\mathrm{GdI_2}$, respectively. The SOC
is incorporated for self-consistent energy and band structure
calculations.
The elastic stiffness tensor  $C_{ij}$   are calculated by using strain-stress relationship (SSR) with GGA,  and  the  piezoelectric stress tensor $e_{ij}$  are carried out by  density functional perturbation theory (DFPT) method\cite{pv6} with GGA.
The 2D elastic coefficients $C^{2D}_{ij}$
 and   piezoelectric stress coefficients $e^{2D}_{ij}$
have been renormalized by   $C^{2D}_{ij}$=$L_z$$C^{3D}_{ij}$ and $e^{2D}_{ij}$=$L_z$$e^{3D}_{ij}$, where the $L_z$ is  the length of unit cell along z direction.  Within finite displacement method, the interatomic force constants (IFCs) of monolayer $\mathrm{GdF_2}$ are calculated based on
the 5$\times$5$\times$1 supercell with FM ground state. Based on the harmonic IFCs, phonon dispersion spectrum of monolayer $\mathrm{GdF_2}$ is obtained by the  Phonopy code\cite{pv5}.

\section{Structure and stability}
The monolayer   $\mathrm{GdCl_2}$ belongs to the hexagonal
crystal system with 2H-$\mathrm{MoS_2}$ type structure, which contains one Gd atomic layer,
sandwiched by two Cl atomic layers (See \autoref{st}).  The corresponding point group is $p\bar{6}m2$ with broken inversion symmetry.
The  magnetic ground state of monolayer   $\mathrm{GdCl_2}$ is determined by comparing the energies of antiferromagnetic (AFM) and FM states with rectangle supercell,  which  is shown in \autoref{energy}. Calculated results show that the FM order is the most stable magnetic state, and the optimized lattice constants with FM order is 3.78  $\mathrm{{\AA}}$, which is consist with
the reported value\cite{g2}. Magnetic anisotropy plays a important role to realize the long-range
magnetic ordering in 2D materials, which can be described by magnetic anisotropy energy (MAE). The monolayer  $\mathrm{GdBr_2}$  and  $\mathrm{GdI_2}$  possess in-plane magnetic anisotropy\cite{g1,g2}. This means that the spin orientations of Gd atoms can be random, and it is difficult to realize the long-range
magnetic ordering without external field. However,  the
easy axis of  monolayer   $\mathrm{GdCl_2}$ is along out of plane direction\cite{g2}.  By considering SOC interaction, the MAE of monolayer $\mathrm{GdCl_2}$ is calculated as the difference between the in-plane  and out-of-plane
magnetization stability energy, and the corresponding value is 93 $\mathrm{\mu eV}$/Gd.

\begin{figure*}
  \includegraphics[width=12cm]{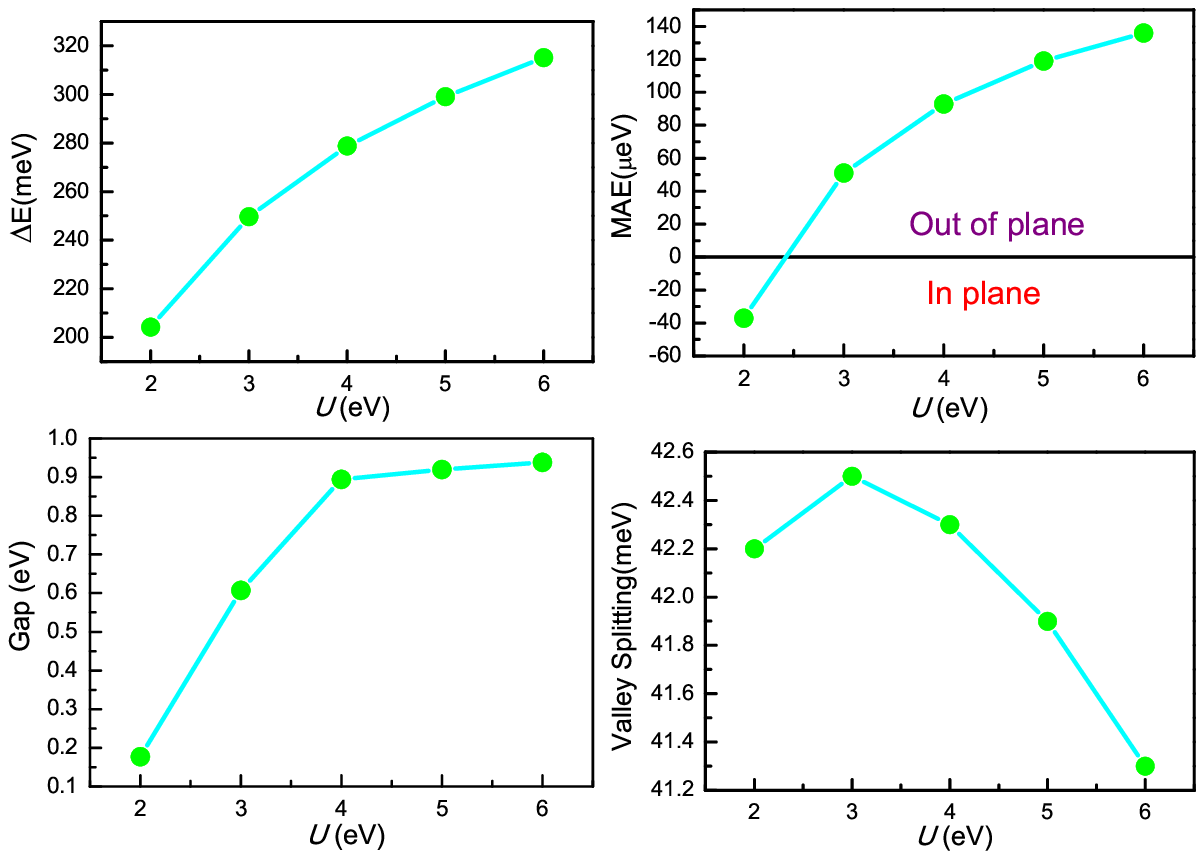}
\caption{(Color online)  (Top left) The energy difference between the AFM and FM states ($\Delta E$); (Top right) MAE; (Bottom left) energy band gap (Gap); (Bottom right) valley splitting
as a function of $U$ for monolayer  $\mathrm{GdCl_2}$.}\label{u-1}
\end{figure*}

\begin{figure}
  \includegraphics[width=8cm]{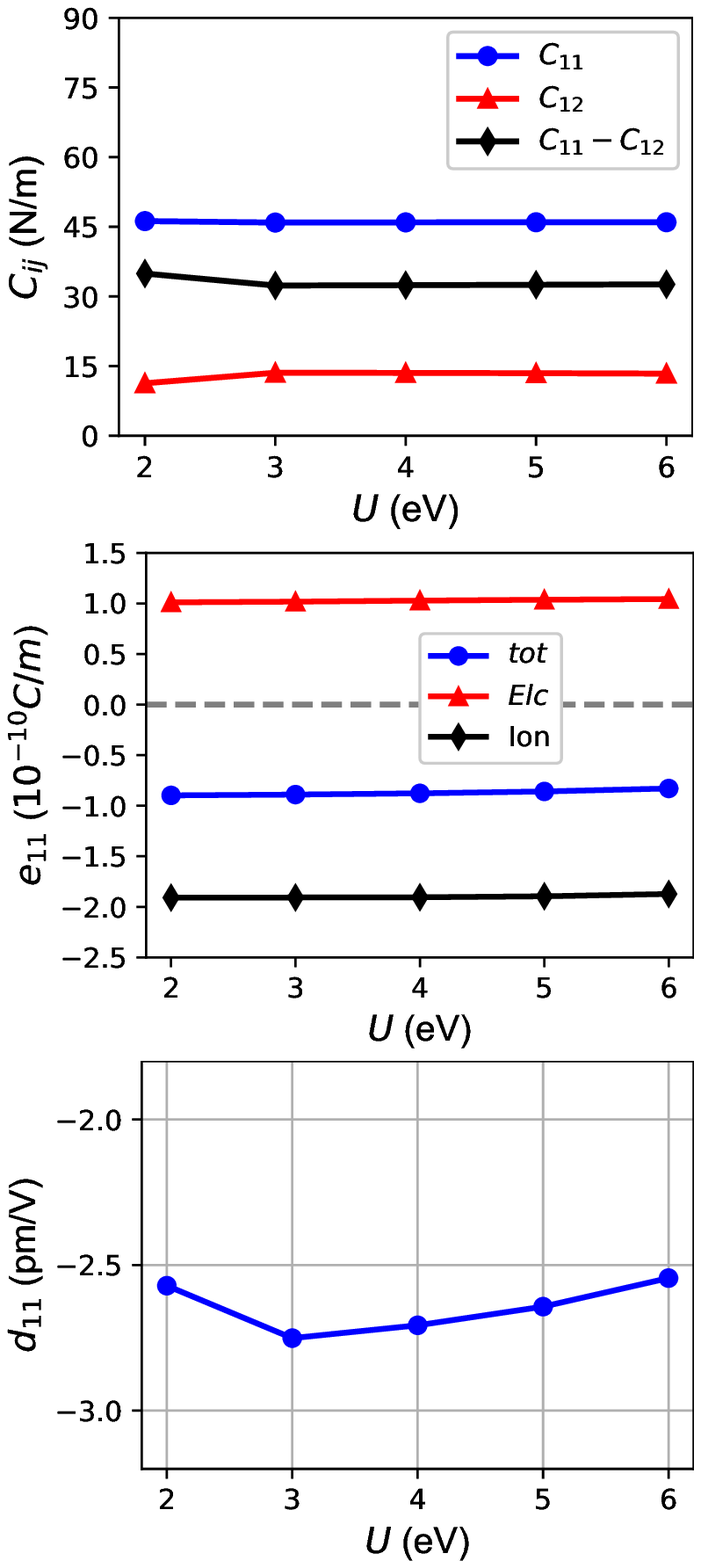}
\caption{(Color online) For monolayer  $\mathrm{GdCl_2}$, the elastic constants  $C_{ij}$,  the piezoelectric stress coefficient  ($e_{11}$) along with the ionic contribution and electronic contribution, and the piezoelectric strain coefficient ($d_{11}$) as a function of $U$.}\label{u-y-d}
\end{figure}

The  thermal and  dynamic stabilities of  monolayer   $\mathrm{GdCl_2}$ have been proved by Ab initio molecular
dynamics (AIMD) simulations and phonon dispersion\cite{g2}. It is also  important to check
the mechanical stability of monolayer   $\mathrm{GdCl_2}$ by  calculating elastic
constants. Using Voigt notation, the elastic tensor $C$ with $p\bar{6}m2$ point-group symmetry for 2D materials can be reduced into:
\begin{equation}\label{pe1-4}
   C=\left(
    \begin{array}{ccc}
      C_{11} & C_{12} & 0 \\
     C_{12} & C_{11} &0 \\
      0 & 0 & (C_{11}-C_{12})/2 \\
    \end{array}
  \right)
\end{equation}
The calculated results show that  $C_{11}$ and $C_{12}$ are 45.95 $\mathrm{Nm^{-1}}$ and 13.53 $\mathrm{Nm^{-1}}$, respectively.  The calculated $C_{ij} $ satisfy the  Born  criteria of mechanical stability\cite{ela}: $C_{11}$$>$0 and $C_{11}-C_{12}$$>$0,  which confirms the mechanical stability of monolayer  $\mathrm{GdCl_2}$. Due to  hexagonal symmetry, the  monolayer  $\mathrm{GdCl_2}$ is  mechanically isotropic. The 2D Young¡¯s moduli $C^{2D}$, shear modulus $G^{2D}$ and Poisson's ratios $\nu^{2D}$ can simply be expressed as\cite{ela}:
\begin{equation}\label{e1}
C^{2D}=\frac{C_{11}^2-C_{12}^2}{C_{11}}
\end{equation}
\begin{equation}\label{e1}
G^{2D}=C_{66}=\frac{C_{11}-C_{12}}{2}
\end{equation}
\begin{equation}\label{e1}
\nu^{2D}=\frac{C_{12}}{C_{11}}
\end{equation}
The calculate  Young's moduli $C_{2D}$,  shear modulus $G_{2D}$ and Poisson's ratio $\nu$  are 41.97  $\mathrm{Nm^{-1}}$, 16.21  $\mathrm{Nm^{-1}}$ and  0.295, respectively.  The  $C_{2D}$ is
less than that of graphene (340 $\mathrm{Nm^{-1}}$)\cite{gra}, which indicates  that  monolayer  $\mathrm{GdCl_2}$ can be easily tuned by strain, being favorable for novel flexible piezotronics.

\section{Electronic structure and valley Hall effect}
The  electronic configuration of  isolated Gd  atom is $4f^75d^16s^2$. For monolayer  $\mathrm{GdCl_2}$,  two electrons of one Gd atom are transferred
to the six neighboring I atoms, and the
electronic configuration of Gd becomes $4f^75d^1$, which will  introduce an  8 $\mu_B$ magnetic moment.
The calculated  magnetic
moment of Gd is 7.463 $\mu_B$, and the total  magnetic moment per unitcell  is 8  $\mu_B$.
The spin-polarized band structure of monolayer  $\mathrm{GdCl_2}$ without SOC is shown in \autoref{band}.
The calculated results show that the monolayer $\mathrm{GdCl_2}$ is a  semiconductors with an indirect band gap of 0.91 eV.
The VBM and CBM are provided by the majority spins and minority spins, and they locate   at the
$\Gamma$ and M high symmetry points, respectively. This makes monolayer  $\mathrm{GdCl_2}$ to be a bipolar magnetic semiconductor, which  can generate 100 \% spin-polarized currents with
inverse spin-polarization direction for electron or hole doping. It is clearly seen that  the energy
extremes of  K and -K high-symmetry points are degenerate in the valence band (\autoref{band} (d)), and monolayer  $\mathrm{GdCl_2}$ is a potential
ferrovalley material.

The band structures of  monolayer  $\mathrm{GdCl_2}$  with SOC for magnetic moment of Gd along the positive and negative z direction (out of plane) are also plotted in \autoref{band}. When the SOC is included, the degeneracy between the K and -K valley states is removed in the valence band, and  a
spontaneous valley polarization is induced with  valley splitting
of 42.3 meV, which is higher than  or compared to ones of reported ferrovalley materials, such as $\mathrm{VAgP_2Se_6}$ (15 meV)\cite{q12}, $\mathrm{LaBr_2}$ (33 meV)\cite{q13,q13-1}, $\mathrm{TiVI_6}$ (22 meV)\cite{q17},  $\mathrm{VSi_2P_4}$ (49.4 meV)\cite{q14} and 2H-$\mathrm{VSe_2}$ (89 meV)\cite{q10}. It is found that the energy of K valley state
is higher than one of -K valley (\autoref{band} (e)).  It is interesting that an external magnetic
field can tune valley polarization of monolayer  $\mathrm{GdCl_2}$. By reversing the magnetization
of Gd atoms,  the spin and valley polarization
can be flipped simultaneously, and the energy of -K valley becomes higher than one  of K valley (\autoref{band} (f)).
These mean that manipulating  magnetization direction is  an efficient way to tune the valley
properties of the monolayer  $\mathrm{GdCl_2}$. Furthermore, the band related with valley  properties is separated well  from other energy bands.
Although the VBM of monolayer  $\mathrm{GdCl_2}$ occurs at $\Gamma$ point, the K/-K valleys are still well defined and not far in energy. In fact, very small compressive strain (about 1\%) can change VBM from $\Gamma$ to  K/-K point (next section). As is well known, the GGA overestimates the lattice constants of materials, and the VBM of  monolayer  $\mathrm{GdCl_2}$ may intrinsically locate at  K/-K point.

The combined effects of the intrinsic
magnetic exchange field and strong SOC give rise to the spontaneous valley polarization.
When the  spin polarization is performed without SOC,
the spin-up and spin-down states are completely split by the
magnetic exchange interaction, but he energy
extremes of  K and -K high-symmetry points are degenerate in the valence band.
When the  magnetic exchange interaction is absent, SOC still can induce
spin nondegeneracy at both K and -K valley  due to missing spatial inversion symmetry,
but K and -K valleys are energetically degenerate with opposite spins because of  existing time reversal
symmetry. In a word,  combined with  high Curie
temperature (224 K)\cite{g2} and PMA,  $\mathrm{GdCl_2}$ is an ideal ferrovalley material for
the  valleytronic devices.

The valley Hall effect can be described by Berry curvature, and  a nonzero Berry curvature along the
out of plane direction  can be attained  in the K and -K valleys for hexagonal systems with broken space inversion symmetry.
With the missing time reversal
symmetry, the valley contrasting feature can be produced. To study these properties  of monolayer  $\mathrm{GdCl_2}$, the Berry curvature
is calculated directly from the calculated
wave functions by using the VASPBERRY code, which is based on Fukui's
method\cite{bm}. The calculated Berry curvature distribution of monolayer  $\mathrm{GdCl_2}$  in the 2D Brillouin zone  without SOC  and  with SOC for magnetic moment of Gd along the positive and negative z direction  are shown \autoref{berry}.
Without SOC,  the  Berry curvatures of K and -K
valleys are  in opposite signs, and the absolute values  are the same.
When the SOC is included, their absolute
values of  the  Berry curvatures of K and -K
valleys are no longer identical,  which shows the typical valley contrasting properties.
It is also found that  the numerical values between K and -K valleys overturn, when  the magnetic moment of Gd changes from  the positive to negative z direction.

\begin{table}
\centering \caption{ For  monolayer  $\mathrm{GdX_2}$ (X=F, Cl, Br and I),   the elastic constants $C_{ij}$ in $\mathrm{Nm^{-1}}$, the piezoelectric stress coefficients $e_{11}$ with electronic part $e_{11e}$ and ionic part $e_{11i}$ in $10^{-10}$ C/m, and the piezoelectric strain coefficients $d_{11}$ in pm/V. }\label{tab}
  \begin{tabular*}{0.48\textwidth}{@{\extracolsep{\fill}}ccccccc}
  \hline\hline
Name& $C_{11}$ &  $C_{12}$&$e_{11e}$& $e_{11i}$& $e_{11}$& $d_{11}$\\\hline\hline
 $\mathrm{GdF_2}$&73.87&19.61&1.365&-1.048&0.317&0.584\\\hline
 $\mathrm{GdCl_2}$&45.95&13.53&1.028&-1.906&-0.878&-2.708\\\hline
 $\mathrm{GdBr_2}$&40.46&11.68&0.847&-1.742&-0.895&-3.110\\\hline
 $\mathrm{GdI_2}$&35.49&10.09&0.658&-1.356&-0.698&-2.748\\\hline\hline
\end{tabular*}
\end{table}

\section{Piezoelectric properties}
The monolayer $\mathrm{GdCl_2}$  with $p\bar{6}m2$  point-group symmetry lacks  inversion symmetry, but
 the reflectional symmetry across the xy plane still holds. These mean that   only  $e_{11}$/$d_{11}$  with  defined x and y direction in \autoref{st} is  nonzero. This is the same with ones of $\mathrm{MoS_2}$ monolayer, but is different from ones of Janus monolayer MoSSe with additional $e_{31}$/$d_{31}$\cite{q7-0}. For 2D materials, only considering  the in-plane strain and stress\cite{q7-0,q7-1,q7-2,q7-3,q7-4,q7-5,q7-6,q7-7,q7-8}, the  piezoelectric stress   and strain tensors by using  Voigt notation  can be reduced into:
  \begin{equation}\label{pe1}
  \left(
    \begin{array}{ccc}
      e_{11} &-e_{11} & 0 \\
    0 &0 & -e_{11}\\
      0 & 0 & 0 \\
    \end{array}
  \right)
  \end{equation}
  \begin{equation}\label{pe1}
  \left(
    \begin{array}{ccc}
        d_{11} & -d_{11} & 0 \\
    0 &0 & -2d_{11} \\
      0 & 0 & 0 \\
    \end{array}
  \right)
  \end{equation}
 When a  uniaxial in-plane strain is imposed,   the in-plane  piezoelectric polarization ($e_{11}$/$d_{11}$$\neq$0) can be induced. However, with an applied   biaxial in-plane strain,  the
in-plane piezoelectric response will be suppressed($e_{11}$/$d_{11}$=0).
The only independent $d_{11}$  can be calculated  by $e_{ik}=d_{ij}C_{jk}$:
\begin{equation}\label{pe2}
    d_{11}=\frac{e_{11}}{C_{11}-C_{12}}
\end{equation}

We use the orthorhombic supercell (in \autoref{st}) to calculate the  $e_{11}$  of monolayer $\mathrm{GdCl_2}$ with DFPT method.
The calculated $e_{11}$ is -0.878$\times$$10^{-10}$ C/m  with ionic part -1.906$\times$$10^{-10}$ C/m  and electronic part 1.028$\times$$10^{-10}$ C/m.
The electronic and ionic polarizations  have  opposite signs, and   the ionic contribution
 dominates the in-plane piezoelectricity.  This is different from  monolayer $\mathrm{MoS_2}$, whose electronic and ionic contributions have the
same sign, and the electronic part dominates the $e_{11}$\cite{q7-8}.
Based on \autoref{pe2}, the $d_{11}$  can be calculated from previous calculated $C_{ij}$ and $e_{11}$.
The calculated  $d_{11}$  is  -2.708 pm/V, which is  comparable to one of  $\alpha$-quartz ($d_{11}$=2.3 pm/V).
The  $\mathrm{GdX_2}$ (X= Br and I) monolayers  have been predicted\cite{g1,g2}, and they possess in-plane magnetic anisotropy.
Here, we use GGA+$U_{eff}$ ($U_{eff}$ = 5.0 and 8.0 eV for monolayer $\mathrm{GdBr_2}$ and $\mathrm{GdI_2}$, respectively) method to  investigate piezoelectric properties of $\mathrm{GdX_2}$ (X= Br and I) monolayers. The data related with elastic and piezoelectric properties are summarized in
\autoref{tab}. It is found that $d_{11}$ of $\mathrm{GdX_2}$ (X= Br and I) monolayers are comparable with one of $\mathrm{GdCl_2}$.

\section{Strain effects}
The VBM of  unstrained monolayer $\mathrm{GdCl_2}$ is at $\Gamma$ point, and it is necessary to tune VBM to K/-K point by external field for practical applications. As is well known,  the  strain is a very effective method to tune the electronic structures of 2D materials\cite{c4,c5,c6,c7,c8,gsd1,gsd2}.
The  $a/a_0$ is used to simulate the biaxial strain with  $a$ and $a_0$ being  the strained and  unstrained lattice constants.
In considered strain range, to confirm the FM ground state,  the energy differences of   AFM  with respect to  FM state  vs $a/a_0$ with rectangle supercell are plotted in \autoref{s-1}.  It is fond the energy difference with the
biaxial strain varying from 0.94 to 1.06 is always positive, and  monotonically decreases.
This indicates that  the ground state of monolayer $\mathrm{GdCl_2}$ is FM in considered strain range, and the
 strain  can strengthen  the FM
coupling between Gd atoms from tensile strain to compressive one.  At applied strain, it is also very important to confirm PMA for
stable  long-range
magnetic ordering without external field. For MAE, \autoref{s-1} shows a decrease with increasing $a/a_0$, and  the MAE becomes negative value with the strain over  1.03, which means that the easy axis of monolayer $\mathrm{GdCl_2}$  turns to in-plane.

We only show energy band structures of monolayer $\mathrm{GdCl_2}$  (0.94 to 1.02) with PMA by using GGA+SOC in \autoref{s-band},
and the energy band gaps  are plotted in \autoref{s-1}.  At applied strain, monolayer $\mathrm{GdCl_2}$ is always an indirect gap semiconductor.
It is found that the compressive strain can induce the transition of VBM from $\Gamma$ point to K/-K point, which can be observed at 0.98 strain.
In fact, the change of VBM has been realized at only 0.99 strain, and the corresponding energy band is plotted in FIG.1 of electronic supplementary information (ESI).
The tensile strain can make CBM change from M point to one point along $\Gamma$-M path. With $a/a_0$ from 0.94 to 1.02, the energy band gap  firstly increases, and then decreases, which can been observed in many 2D materials\cite{gsd1,gsd2}.
As shown in \autoref{s-1},
the valley splitting increases monotonically with the increasing $a/a_0$.  Conversely, a compressive strain decreases
the valley splitting, and  the valley splitting will become negative value at about 0.963 strain,
which implies that the energy of -K valley
is higher than one of K valley. The Berry curvature distributions of monolayer  $\mathrm{GdCl_2}$   with $a/a_0$ being 0.94, 0.98 and 1.02 by using GGA+SOC are shown in \autoref{s-berry}.  It is found that the Berry curvatures  (absolute value) of two
valleys become large with increasing $a/a_0$.

 It have been proved that strain engineering can  effectively tune   piezoelectric properties of 2D materials\cite{r1,r2,r3,r4}, and then  we investigate  the strain effects on piezoelectric properties of monolayer  $\mathrm{GdCl_2}$.
 The elastic constants  ($C_{11}$, $C_{12}$ and $C_{11}$-$C_{12}$),  piezoelectric  stress  coefficients  ($e_{11}$) along  the ionic  and electronic contributions,  and  piezoelectric  strain  coefficients ($d_{11}$) of monolayer  $\mathrm{GdCl_2}$ as a function of $a/a_0$ are plotted in \autoref{y-d}. It is clearly seen that $C_{11}$, $C_{12}$ and $C_{11}$-$C_{12}$   all decrease with increasing strain from 0.94 to 1.02 strain.
In the considered strain range, the calculated elastic constants of strained  monolayer  $\mathrm{GdCl_2}$ satisfy  the mechanical stability criteria\cite{ela}, so they are all  mechanically stable.
It is found that the  strain has little  effects on $e_{11}$,  including both the ionic  and electronic contributions.
However, with  increasing $a/a_0$, the $d_{11}$ (absolute value) increases due to reduced $C_{11}$-$C_{12}$ based on \autoref{pe2}.

Considering various factors, very small  compressive strain (about 0.99 strain) can make monolayer  $\mathrm{GdCl_2}$ to be a good valley material with
PMA, VBM at K/-K point, strong FM coupling, proper valley splitting and $d_{11}$ to realize PAVHE.

\section{electronic correlation effects}
To further confirm the reliability of our results, the  electronic correlation effects on magnetic, electronic and piezoelectric properties of monolayer  $\mathrm{GdCl_2}$ are
investigated by choosing different $U$ (2-6 eV).  The energy differences between   AFM  and FM states   with rectangle supercell and MAE  vs $U$ are plotted in \autoref{u-1}. With increasing $U$,  the energy difference is always positive, and  monotonically increases.
These manifest that the monolayer $\mathrm{GdCl_2}$ is always  FM  order, and the increasing $U$ can strengthen  the FM
coupling between Gd atoms.
The MAE shows a decrease with decreasing $U$, and  the MAE becomes negative value with easy axis turning to in-plane
with $U$ being less than about 2.5 eV.
The  energy band structures of monolayer $\mathrm{GdCl_2}$  ($U$=2 to 6 eV)  by using GGA+SOC  are plotted in FIG.2 of ESI,
and the energy band gaps and  valley splitting are plotted in \autoref{u-1}.
When the $U$ increases,  the VBM is always at K/-K point, and the gap increases. It is found that  the electronic correlation has little influence on valley splitting, and the change only 1.2 meV with different $U$ (2-6 eV). From FIG.3 of ESI, the electronic correlation has little effects on
Berry curvatures of K and -K
valleys.
The elastic constants  ($C_{11}$, $C_{12}$ and $C_{11}$-$C_{12}$),  piezoelectric  stress  coefficients  ($e_{11}$) along  the ionic  and electronic contributions,  and  piezoelectric  strain  coefficients ($d_{11}$) of monolayer  $\mathrm{GdCl_2}$  vs $U$ are shown in \autoref{u-y-d}.
Calculated results show that electronic correlation has small effects on $d_{11}$ due to small influence on $C_{ij}$ and $e_{11}$, and the change is about 0.21 pm/V.

\begin{figure}
  \includegraphics[width=8cm]{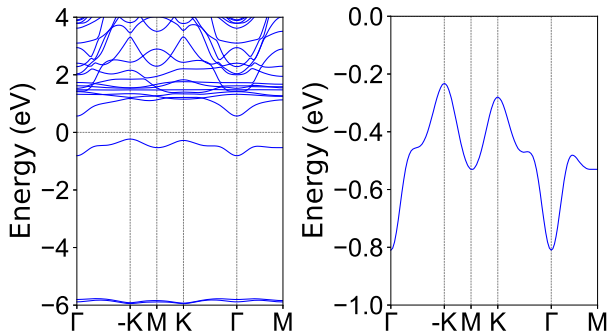}
  \caption{(Color online) The band structure of  monolayer  $\mathrm{GdF_2}$  with SOC for magnetic moment of Gd along the positive  z direction (Left) with enlarged views of the valence bands near the Fermi level (Right).}\label{band-F}
\end{figure}

\section{Discussion and Conclusion}
The  $\mathrm{GdX_2}$ (X=Cl, Br and I) monolayers have been predicted\cite{g1,g2}, and the easy axis of monolayer $\mathrm{GdCl_2}$ is along the out of plane direction,
while monolayer $\mathrm{GdBr_2}$ and $\mathrm{GdI_2}$ possess in-plane magnetic anisotropy. When the compressive strain is larger than 3\%, the easy
axis of monolayer $\mathrm{GdBr_2}$ transfers from in-plane to out-of-plane\cite{g2}. These mean that monolayer $\mathrm{GdF_2}$ should have PMA due to small atomic radius of F atoms.
For monolayer $\mathrm{GdF_2}$,  the energy difference between AFM and FM is 0.285 eV, which means that the FM order  is the ground state.
The optimized lattice constants is 3.465 $\mathrm{{\AA}}$,  and the calculated $C_{11}$ and $C_{12}$ are  73.87 $\mathrm{Nm^{-1}}$ and  19.61 $\mathrm{Nm^{-1}}$,  satisfying the  Born  criteria of mechanical stability\cite{ela}.
From FIG.4 of ESI, the  monolayer $\mathrm{GdF_2}$ is dynamically stable due to missing imaginary frequency.
The PMA of monolayer  $\mathrm{GdF_2}$  is confirmed, and the corresponding MAE is  137 $\mu$eV per Gd atom.
The band structures of  monolayer  $\mathrm{GdF_2}$  with SOC for magnetic moment of Gd along the positive z direction  are  plotted in \autoref{band-F}. The monolayer  $\mathrm{GdF_2}$ is a indirect gap semiconductor (0.80 eV) with VBM at K/-K point and CBM at $\Gamma$ point.
It is found that the energy of -K valley
is higher than one of K valley with valley splitting of 47.6 meV.
Calculated Berry curvature distribution of monolayer  $\mathrm{GdF_2}$  with SOC for magnetic moment of Gd along the positive  z direction is plotted in FIG.5 of ESI.  The absolute
values of  the  Berry curvatures of K and -K
valleys are smaller than ones of monolayer  $\mathrm{GdCl_2}$, and are  no longer identical with the typical valley contrasting properties.
Finally, the piezoelectric properties of monolayer $\mathrm{GdF_2}$ are investigated, and  the calculated   $d_{11}$ is   0.584 pm/V.
With respect to $\mathrm{GdX_2}$ (X=Cl, Br and I) monolayers,  the $d_{11}$ of monolayer  $\mathrm{GdF_2}$ becomes positive, which is because the electronic part of monolayer  $\mathrm{GdF_2}$ is larger than ionic one.
The related data are summarized in \autoref{tab}. These results show that monolayer  $\mathrm{GdF_2}$  may be a potential valley material to achieve PAVHE.

In summary,  a possible way is proposed  to achieve anomalous valley Hall effect  by piezoelectric effect, and then the  valleytronic and piezoelectric properties of
 monolayer  $\mathrm{GdCl_2}$ are investigated by the reliable first-principle calculations.  Monolayer  $\mathrm{GdCl_2}$  is a FM
semiconductor with a pair of valleys locating at the K and -K
points, and possess PMA. Arising from the intrinsic   magnetic
interaction, broken inversion symmetry and SOC, the valley splitting
can be observed  between K and -K valleys, and the corresponding value is 42.3 meV. The predicted $d_{11}$ is -2.708 pm/V, which can provide
a suitable in-plane electric field by  uniaxial in-plane strain.  Moreover, the effects of strain and electronic correlation on
the valley physics and piezoelectric properties are also studied.
Finally, a 2D  FM semiconductor $\mathrm{GdF_2}$  is predicted, which is also a potential ferrovalley material.
 Our work provides an initial idea  for realizing and
manipulating the valley  physics.

\begin{acknowledgments}
This work is supported by Natural Science Basis Research Plan in Shaanxi Province of China  (2021JM-456). We are grateful to the Advanced Analysis and Computation Center of China University of Mining and Technology (CUMT) for the award of CPU hours and WIEN2k/VASP software to accomplish this work.
\end{acknowledgments}

\end{document}